\documentclass[onecollarge]{svjour2}
\usepackage{epsfig}
\usepackage{amsmath}
\usepackage{amssymb}
\usepackage{graphicx}
\usepackage{color}
\usepackage{epstopdf}
\journalname{Few-Body Syst}

\begin{document}

\title{Efimov Trimers near the Zero-crossing of a Feshbach Resonance}

\author{N.~T. Zinner}
\institute{ Department of Physics and Astronomy,
         Aarhus University, DK-8000 Aarhus C, Denmark }

\date{\today}

\maketitle

\begin{abstract}
Near a Feshbach resonance, the two-body scattering length can assume any value. When it approaches zero, the
next-order term given by the effective range is known to diverge. We consider the question of whether this
divergence (and the vanishing of the scattering length) is accompanied by an anomalous solution of the 
three-boson Schr{\"o}dinger equation similar to the one found at infinite scattering length by Efimov. 
Within a simple zero-range model, we find no such solutions, and conclude that higher-order terms
do not support Efimov physics.
\end{abstract}

\section{Introduction}
In a three-body system of identical bosons, an infinite sequence 
of bound states appears at the threshold for binding of 
the two-body subsystems as shown originally by Efimov in the 
context of nuclear physics \cite{efimov1970,efimov1971}. However,
such states unfortunately do not exist in nuclear system 
\cite{jensen2004}.
Efimov states have been a focus of the ultracold gas community 
since the experimental observations of signatures of such states in 
a dilute gas of Cesium atoms \cite{grimm2006,grimm2009,grimm2011}. 
This was followed by studies in three-component 
$^{6}$Li \cite{otten2008,huckans2009,wenz2009,naka2010,naka2011},
bosonic $^{7}$Li \cite{gross2009,gross2010} and $^{41}$K \cite{zaccanti2009}, 
and heteronuclear mixtures of $^{41}$K and $^{87}$Rb \cite{barontini2009}. 
By sweep a Feshbach resonance in $^{7}$Li over several orders of magnitude 
\cite{pollack2009} a recent study \cite{hulet2009} was able to 
identify eleven distinct features in the recombination rates. 
Some of these are interpreted as four-body resonances that have been theoretically
predicted as associated to the Efimov 
trimers \cite{platter2004,hanna2006,vstecher2009}.
Evidence for potential four-body resonances was also
found in \cite{zaccanti2009,ferlaino2009}. The universal theory 
of four-body states has, however, been questioned \cite{yamashita2006}, and it is 
unlikely that universality persists to higher particle numbers \cite{yamashita2009}.

The characteristic geometric energy and size scaling of Efimov states 
near a resonance has been discussed within various models by 
many 
people (see \cite{nielsen2001} and \cite{braaten2006} for comprehensive reviews). 
In cold gases one can use Feshbach resonances to tune interactions
\cite{timmermans1999,bloch2008,giorgini2008,chin10} and thus reach resonance 
conditions in a controlled way. This gives the celebrated 
scaling factor $e^{-\pi n/s_0}\approx 22.7$ with $s_0=1.00624$. 
The energy of the $n$th state is then simply related 
to the first one by $E(n)/E(0)=22.7^{-2n}$ and the root-mean-square size 
by $r_{rms}(n)/r_{rms}(0)=22.7^{-n}$. This result
is found within the universal theory where the scattering length is 
much larger than the range of the interatomic potential. 
However, evidence for systematic deviations from the universal 
scaling has been found \cite{zaccanti2009}, and therefore
it is necessary to consider correction to the universal picture 
in detail. The experiment reported in Ref.~\cite{zaccanti2009} was found to 
agree qualitatively with the corrections found in \cite{tfj08c,pjp09}.

In this paper we study the corrections to Efimov physics as 
the width of the Feshbach resonance becomes smaller. In
this case the effective-range is sizable and must be 
taken into account. The zero-range model is attractive due to 
its simplicity and range corrections have been discussed 
in Refs.~\cite{pjp09,fj01b,fj02,jonsell2004,tfj09,wang11}. 
Here we use
a two-channel model to describe the Feshbach resonance \cite{bruun2005}. 
This kind of model has been considered by several authors 
for both broad and narrow resonances in few- and many-body contexts
\cite{phil98,kokkelmans02,petrov03,jonsell06,mas06,diener06,pet07,braaten08,gogolin08,mas08,plat09,pri09,pri10,pri11,castin11,naidon11}.
A two-channel model gives the energy dependence of 
the interaction in terms of the background scattering length and 
the width of the resonance. This is essential 
since we will be interested in the behavior of the 
corrections both around the resonance and around the 
point where the scattering length goes to zero. We recently 
considered the effects of Feshbach resonances around zero-crossing
for a macroscopic Bose-Einstein 
condensate \cite{zinner2009a,thoger2009} and found considerable
changes in critical particle numbers for very narrow resonances. 

Here we consider the interesting question of whether higher 
order corrections tied to the effective-range term can 
produce anomalous solutions to the three-boson problem 
as seen around the resonance. While previous studies using
finite-range models \cite{tfj09} and full numerical three-body
calculations \cite{wang11} have considered similar situations, 
we simplify the discussion considerably by using zero-range model with a 
two-channel description of the scattering process. This 
recovers known results about both broad and narrow Feshbach
resonances, i.e. the Efimov effect occurs at the point where the 
scattering length diverges. In addition, we investigate the solutions
around the point where the scattering length goes to 
zero. Within the two-channel zero-range model, the effective 
range diverges when the scattering length is zero, and 
one may wonder whether this divergence also leads to interesting
three-body phenomena. We find no evidence for anomalous solutions
that could produce a spectrum similar to the one found by 
Efimov at infinite scattering length, regardless of whether the resonance is broad or
narrow.

\section{Feshbach Model}
Since we are interested in finite-range corrections within the zero-range 
approximation, we need to consider multi-channel models. In the case of 
$s$-waves, two-channel models have been derived using various methods, 
including low-energy effective theory \cite{bruun2005}, 
multi-channel quantum-defect theory \cite{mies84,julienne06}, 
and resonance models \cite{kokkelmans02}. The on-shell open-open 
channel $s$-wave $T$-matrix as a function of magnetic 
field strength, $B$, can be written
\begin{align}
T(k)=\frac{\frac{4\pi\hbar^2a_{bg}}{m}}{\left(1+\frac{\Delta\mu\Delta B}{\frac{\hbar^2k^2}{m}-\Delta\mu(B-B_0)}\right)^{-1}+ia_{bg}k},
\end{align}
where $m$ is the atomic mass, $k$ is the center-of-mass momentum, 
$a_{bg}$ is the background $s$-wave scattering length, $B_0$ is the 
position of the resonance, $\Delta B$ is the resonance width, and 
$\Delta\mu$ is the difference in magnetic moment of the open and 
closed channels. We can compare this to the usual expression for the 
$T$-matrix in vacuum 
\begin{align}
T(k)=-\frac{\frac{4\pi\hbar^2}{m}}{k\cot\delta(k)-ik}
\end{align}
to get an expression for $k\cot\delta(k)$ as a function of $B$. This yields
\begin{align}
k\cot\delta(k)=-\frac{1}{a_{bg}}\left(1+\frac{\Delta\mu\Delta B}{\frac{\hbar^2k^2}{m}-\Delta\mu(B-B_0)}\right)^{-1}.
\end{align}
If we neglect the $k$-dependent term we recover the usual formula 
\begin{align}
&k\cot\delta(k)=-\frac{1}{a(B)}&\\
&a(B)=a_{bg}\left(1-\frac{\Delta B}{B-B_0}\right).&
\end{align}
Keeping the $k$ term and going to the resonance where $B=B_0$ we find
\begin{align}
k\cot\delta(k)=-\frac{\frac{1}{2}\frac{2\hbar^2}{m\Delta\mu\Delta B a_{bg}}k^2}{1+2\frac{\hbar^2}{2m\Delta\mu\Delta B a_{bg}}a_{bg}k^2}.
\end{align}
If we define $r_{e0}=-2\hbar^2/m a_{bg}\Delta\mu\Delta B$ and make 
a low momentum expansion we see that this is consistent with an 
effective range expansion where the constant term is absent as 
the $a(B)$ diverges at $B=B_0$. Notice that $a_{bg}\Delta\mu\Delta B>0$ 
for all Feshbach resonances \cite{chin10} and therefore $r_{e0}<0$ 
always. Similarly, at zero-crossing where $B=B_0+\Delta B$, we find
\begin{align}
k\cot\delta(k)=-\frac{1}{a_{bg}}-\frac{2}{a_{bg}^{2}r_{e0}}k^{-2}.
\end{align}
We thus see that no meaningful effective range expansion is possible. 
However, the $T$-matrix stays finite and one can define an effective 
potential at zero-crossing that involves $a_{bg}$ and $r_{e0}$ \cite{zinner2009c}. 
If we introduce the function $f(B)=(B-B_0)/\Delta B$ we can  write the 
general form
\begin{align}\label{phase}
k\cot\delta(k)=\frac{1}{a_{bg}}\left[\frac{f(B)-\tfrac{1}{2}a_{bg}|r_{e0}|k^2}{1-f(B)+\tfrac{1}{2}a_{bg}|r_{e0}|k^2}\right],
\end{align}
where $r_{e0}<0$ was used. We recover the expression above at 
resonance where $f(B_0)=0$ and zero-crossing where $f(B_0+\Delta B)=1$. An expansion of the phase-shift
to second order gives an expression for the magnetic field dependent effective range of the 
form \cite{zinner2009c}
\begin{align}
r_e(B)=r_{e0}\left[1-\frac{a_{bg}}{a(B)}\right]^2,
\end{align}
which clearly shows the divergent nature of $r_e(B)$ when $a(B)\to 0$. Higher-order corrections do, 
however, depend on the combination $a^2 r_e$ which has a finite limit \cite{zinner2009a,thoger2009}.

\section{Effective Hyperspherical Potential}
We will employ the hyperspherical formalism \cite{nielsen2001,braaten2006,fedorov1993}, 
where Efimov states are found from an eigenvalue equation. 
In the universal limit where only the scattering length 
is important we have
\begin{align}\label{eig}
\frac{-\nu\cos(\nu\tfrac{\pi}{2})+\tfrac{8}{\sqrt{3}}\sin(\nu\tfrac{\pi}{6})}{\sin(\nu\tfrac{\pi}{2})}=-\frac{\sqrt{2}\rho}{a}.
\end{align}
Here $\rho$ is the hyperspherical radius defined for three 
identical particles by $\rho^2=\tfrac{1}{3}(r_{12}^{2}+r_{13}^{2}+r_{23}^{2})$ 
where $r_{ij}$ is the distance between particles $i$ and $j$ \cite{nielsen2001}. 
For later convenience, we define the function $g(\nu)$ to be the left-hand side
of Eq.~\eqref{eig}, and also
\begin{align}\label{gs}
g(s)=\frac{-s\cosh(s\tfrac{\pi}{2})+\tfrac{8}{\sqrt{3}}\sinh(s\tfrac{\pi}{6})}{\sinh(s\tfrac{\pi}{2})},
\end{align}
which is obtained from the left-hand side of Eq.~\eqref{eig} by the substitution
$\nu=is$.
The solution of this equation for each value of $\rho$ gives the effective 
hyperradial potential
\begin{align}\label{veff}
V_{eff}(\rho)=\frac{\hbar^2}{2m}\frac{\nu(\rho)^2-\tfrac{1}{4}}{\rho^2}-Q(\rho),
\end{align}
where $Q(\rho)$ is a diagonal coupling term in the hyperspherical 
formalism \cite{nielsen2001}.
At the resonance where $|a|=\infty$, we find a constant solution $\nu=i s_0$ where $s_0=1.00624$ 
(which gives the scaling parameter discussed in the introduction). In this 
case the hyperradial equation for large distances reduces to 
\begin{align}\label{radial}
\left(-\frac{\partial^2}{\partial\rho^2}- \frac{s_{0}^{2}+\tfrac{1}{4}}{\rho^2}+\kappa^2\right)f_0(\rho)=0,
\end{align}
where $-\hbar^2\kappa/2m=E$ is the energy of the three-body state. 
Here we have neglected $Q(\rho)$ at large distance; 
i.e. $r_0\ll \rho\ll |a|$ \cite{nielsen2001}, where $r_0$ is 
a small distance cut-off (sometimes refered to as a three-body 
parameter). In the rest of this paper we will
be interested in the asymptotic region $r_0\ll \rho$. 
When there
is a two-body bound in the system (as we will assume below), 
the leading order in $Q(\rho)$ behaves as $-1/4\rho^2$ at 
short distance, while in the absence of such two-body bound state,
$Q(\rho)$ falls off faster than $1/\rho^2$ \cite{fj01b,nielsen1999}.
In any case, we are interested in the occurence of the
Efimov effect which hinges on the coefficient of the $1/\rho^2$
in the region $r_0\ll \rho$, so we neglect the contribution from 
$Q(\rho)$ from now on.

Introducing scaled variables $\tilde f_0=f_0/\sqrt{\rho}$ and 
$\tilde\rho=\kappa\rho$, Eq.~\eqref{radial} becomes a Bessel equation. The 
solution is $\sqrt{\rho}K_{is_0}(\kappa\rho)$ where $K$ is the modified
Bessel function of the second kind. As the energy goes to zero 
we recover the Efimov spectrum which arises because $K_{is_0}(\rho)$ 
is log-periodic at low energies and thus allows for infinitely
many nodes \cite{nielsen2001}. Below we will be concerned with 
changes in the Efimov
scaling as we vary the parameters of the two-channel Feshbach model.

In the above we have neglected 
couplings to higher partial waves in the adiabatic expansion, 
which are generally smaller than the leading term considered here. 
Higher adiabatic components come with larger centrifugal barriers
and it is therefore much more difficult to obtain attraction is these
channels. It is therefore natural to look for anomalous solutions in 
the lowest component studied here.

Two-channel models can also be studied in
momentum-space using both zero-range \cite{pri10} and small finite-range
implementations \cite{pri11} that are essentially exact. However, this
is somewhat more complicated than our approach of adiabatic expansion 
and retention of only the lowest order radial term. While this has
mostly been studied in the case of large scattering length, we note
that the phase-shift given in Eq.~\eqref{phase} is well-behaved also
for small scattering length \cite{zinner2009c} and it can therefore
serve as a boundary condition for the adiabatic approximation as 
discussed below. While there will in general be couplings to higher
adiabatic components (even in the limit of diverging scattering length), 
we do not expect this to influence the presence or absence of 
anomalous solutions in the lowest component.

We also assume that the well-known Thomas collapse is remedied by a 
boundary condition 
at small hyperradius; $f_0(r_0)=0$. The latter condition in fact 
determines the position of the first Efimov state and calibrates the 
spectrum. It is not clear whether the actual value of $r_0$ can 
be determined 
in the universal theory or not. Recent measurements seem to suggest
that $r_0$ is simply proportional to $r_\textrm{vdW}$ \cite{benig2011}
but the reasons are not entirely understood (see Ref.~\cite{chin11}
for a simple argument as to why this should be true).

The next step is to replace the scattering length in Eq.~\eqref{eig} 
by the full phase shift, which means changing the 
boundary condition from $-1/a$ to $k\cot\delta(k)$. 
In order to do this we have to use the 
correspondence between $k$ and the hyperradius $\rho$ which is 
$k\rightarrow k_\rho=\nu(\rho)/(\sqrt{2}\rho)$ \cite{fj01b,tfj09}. 
We can then put 
$k_\rho\cot\delta(k_\rho)$ on the right-hand side of Eq.~\eqref{eig}, 
where $\delta(k_\rho)$ is the phase-shift of the scattering of a 
pure two-particle system evaluated at momentum $k_\rho$. This approach 
was already used in \cite{fj01b,jonsell2004}. However, as discussed 
in Ref.~\cite{tfj09}, it is in fact an approximation. The potential one 
should use instead is the modified two-body term in 
the hyperspherical formalism 
$V_{\rho}(r)=V(\sqrt{2}\rho\sin(\tfrac{r}{\sqrt{2}\rho}))$. As was 
shown in \cite{tfj09}, for an effective range expansion this gives 
corrections to the effective potential as compared to using the 
standard two-body phase shift and the result in fact deviates from 
the predictions of effective field theory \cite{pjp09}. The corrections 
to the effective potential are of order $1/\rho^3$ and in the large 
distance regime that we are currently interested in we will neglect 
this and use the simple approximation with $\delta(k_\rho)$. We 
expect that an analysis along the lines of \cite{tfj09} can be made 
for the two-channel model as well and it would be interesting to 
see the quantititative nature of this in the future.

Replacing $-1/a$ in Eq.~\ref{eig} by Eq.~\eqref{phase} with $k_\rho=\nu/(\sqrt{2}\rho)$, we find
\begin{align}
g(\nu)=\frac{\sqrt{2}\rho}{a_{bg}}\left[\frac{f(B)
-\tfrac{1}{4}a_{bg}|r_{e0}|\tfrac{\nu^2}{\rho^2}}{1-f(B)+\tfrac{1}{4}a_{bg}|r_{e0}|\tfrac{\nu^2}{\rho^2}}\right].
\label{final}
\end{align}
We will now proceed to solve this equation for $\nu(\rho)$ and 
compare to the universal case for different values of $B$. The 
universal Efimov
scaling of energies and radius emerges in the limit of constant $\nu=is_0$. 
When the potential deviates from the constant value it is relevant
to ask over what region (if any) $\nu$ remains essentially flat and what 
the numerical value is that gives the scaling. The number of bound states 
is then roughly proportional to $\ln(\rho_1/\rho_0)$ where $\rho_0$ and 
$\rho_1$ are then boundaries of the flat region 
\cite{efimov1970,efimov1971,nielsen2001}.

\section{Analysis}
We will consider the case of $a_{bg}>0$ only, the opposite case is 
similar. This implies that there is a two-body bound state in the 
scattering channel. This is reflected in the hyperradial potential
for $\rho<a_{bg}$ as discussed in Ref.~\cite{fj01b}. Since we are 
interested in three-body physics only, we will consider the region
$\rho>a_{bg}$ only. Having introduced the function $f(B)$ above, 
we can split the analysis of Eq.~\eqref{final} into two cases, i) 
near resonance, $f(B)=0$, and ii) near zero-crossing, $f(B)=1$. 
We proceed to analyse these in turn.

\subsection{Near resonance}
Here we have $f(B)\to 0$. The right-hand side of Eq.~\eqref{final}
becomes
\begin{align}\label{res}
\frac{\sqrt{2}\rho}{a_{bg}}\left[\frac{-\tfrac{1}{4}\tfrac{|r_{e0}|}{a_{bg}}\nu^2\left(\tfrac{a_{bg}}{\rho}\right)^2}
{1+\tfrac{1}{4}\tfrac{|r_{e0}|}{a_{bg}}\nu^2\left(\tfrac{a_{bg}}{\rho}\right)^2}\right].
\end{align}
First consider the case $a_{bg}\gg|r_{e0}|$. Then for any finite $\rho$, Eq.~\eqref{res}
will go to zero. This will also be true for the case where $\rho\gg a_{bg}$. Either
way we have $g(\nu)=0$ which is known to have the anomalous solution $\nu=is_0$
as discussed above. We thus have Efimov trimers in the region bounded by
$a_{bg}\ll \rho\ll |a|$ as usual. For shorter distances, we have two-body 
physics at play. In the case $a_{bg}\sim |r_{e0}|$, the conclusion is 
the same.

Next consider the case where $|r_{e0}|\gg a_{bg}$ which is the case for 
narrow Feshbach resonances \cite{wang11,chin10}. For any finite $\rho<|r_{e0}|$, 
we find instead of Eq.~\eqref{res}, that the right-hand side of Eq.~\eqref{final}
will behave like $-\sqrt{2}\rho/a_{bg}$. Looking at Eq.~\eqref{gs}, we see
that $g(s)$ becomes a linear function of $s$ for $s\gg 1$. A solution of the 
form $\nu(\rho)=\alpha i \rho/a_{bg}$ can therefore be found in the region
$a_{bg}\ll \rho<|r_{e0}|$. However, the effective potential in Eq.~\eqref{veff}
then gets a constant contribution and this will not produce an Efimov-like 
spectrum since the requirement is that $\nu^2$ be constant and negative over
an extended region.

The final region to consider is $\rho\gg |r_{e0}|$. Here we see from 
Eq.~\eqref{res} that we recover $g(\nu)=0$ and the condition for the 
Efimov effect is fulfilled. The region is $|r_{e0}|\ll \rho<|a|$ 
as previously found \cite{fj01b,tfj09,wang11}. The zero-range model 
is thus seen to reproduce this intuitively clear result in a simple 
and elegant manner.

\subsection{Near Zero-crossing}
Next we consider the case of zero-crossing where $a\to 0$, $f(B)\to 1$, 
and the right-hand side of Eq.~\eqref{final} becomes
\begin{align}\label{zero}
4\sqrt{2}\left(\frac{\rho}{a_{bg}}\right)^3\left[\frac{1-\tfrac{1}{4}\tfrac{|r_{e0}|}{a_{bg}}\nu^2\left(\tfrac{a_{bg}}{\rho}\right)^2}
{\tfrac{|r_{e0}|}{a_{bg}}\nu^2}\right].
\end{align}
First consider the case where $a_{bg}\gg |r_{e0}|$. In this case, Eq.~\eqref{zero}
blows up because of the $|r_{e0}|/a_{bg}$ term in the denominator. Neglecting the 
second term in the numerator (which is small since we are still assuming 
that $\rho\gg a_{bg}$), we can multiply Eq.~\eqref{final} by $\nu^2$ and 
in turn repeat the argument above about the behavior of $g(s)$ for large 
values of $s$ to obtain a solution that goes like $\nu\propto i\beta\rho/a_{bg}$
where $\beta$ contains a factor $(a_{bg}/|r_{e0}|)^{1/3}$ which we assumed to 
be large. Again we are lead to the conclusion that the effective hyperradial 
potential in Eq.~\eqref{veff} will only receive a constant contribution in 
the region $|r_{e0}|\ll a_{bg}\ll \rho\ll |f(B)-1|^{-1}$. The upper bound
is given by how close we are to the point where $a$ vanishes, similar to the 
case near resonance where $a$ is the upper bound. At zero-crossing, the 
interval becomes unbounded from above, and no Efimov-like potential will
emerge anywhere for resonances where $a_{bg}\gg |r_{e0}|$. The same 
is true for $a_{bg}\sim |r_{e0}|$ by similar analysis.

Now consider instead the opposite limit $|r_{e0}|\gg a_{bg}$. For
$\rho\ll|r_{e0}|$, Eq.~\eqref{zero} becomes simply $-\sqrt{2}\rho/a_{bg}$.
The solution in this case is again $\nu\propto i\gamma\rho/a_{bg}$ and 
nothing interesting happens. The last regime where one can still hope
for a non-trivial solution is when $\rho\gg|r_{e0}|$, which is where 
Efimov trimers appear near resonance. However, in similar fashion 
to the case where $a_{bg}\gg|r_{e0}|$, we find that $\nu\propto i\delta\rho/a_{bg}$
with $\delta$ containing a factor $(a_{bg}/|r_{e0}|)^{1/3}$ which is now 
a small number. This does not, however, change the outcome and no 
Efimov-like solution is found.

We have done numerical root finding on Eq.~\eqref{final} to confirm the analysis
above. We have tested the numerics and made sure that it reproduces the 
usual Efimov solution, $s_0$, and also that it finds the free solutions \cite{nielsen2001}.
In the case of zero-crossing, we have not found any constant and 
purely imaginary solution for $\nu$.
Thus, numerically we also do not see
any signs that Efimov-like features are present around zero scattering length.

\section{Summary}
We have considered the occurence of a geometric spectrum in the 
three-boson system around Feshbach resonances in ultracold gases 
when effective-range corrections are included. In particular, we
have explored the possibility of having an Efimov-like effect 
around the zero-crossing where the scattering length goes to 
zero and higher order effective-range terms must be considered. 
This is done within a simple zero-range model that allows one
to obtain a closed equation for the coefficient of the effective
potential in the hyperspherical approach. The effective potential 
in turn determines whether there can be universal three-body states
with geometric scaling properties.

In the analysis of the closed equation, we recover the conclusions of 
previous studies, including those obtained through numerical 
approaches. Close to the resonance, the region of space where 
universal trimers are supported is bounded from below by either
the background scattering length or the background effective range, 
whichever is larger, while it is bounded from above by the value
of the scattering length (which goes to infinite on resonance).

Around the point where the scattering length crosses zero, the 
zero-range model predicts that no region can be found where the 
effective potential can support Efimov trimer states. This 
is true for both broad and narrow Feshbach resonances.
Thus even though the magnetic field dependent effective range
of a two-channel model diverges around zero-crossing, this
does not seem to support a geometric spectrum of Efimov trimers.
More generally, this implies that higher-order terms do not 
add attraction in the three-body channel of a many-body system.
This is consistent with the fact that the instabilities observed 
in condensates with higher-order interaction terms 
in Refs.~\cite{zinner2009a,thoger2009} originate from two-body 
effects.

\end{document}